\documentstyle[11pt]{article}                                   

\makeatletter
\def\section{\@startsection {section}{1}{\z@}{-3.5ex plus -1ex minus
 -.2ex}{2.3ex plus .2ex}{\large\bf}}
\def\subsection{\@startsection{subsection}{2}{\z@}{-3.25ex plus -1ex minus
 -.2ex}{1.5ex plus .2ex}{\normalsize\bf}}
\makeatother
\makeatletter
\@addtoreset{equation}{section}

\makeatother
\newcommand{\nc}{\newcommand}
\newcommand{\rnc}{\renewcommand}
\nc{\bea}{\begin{eqnarray}}
\nc{\eea}{\end{eqnarray}}
\nc{\be}{\bea}
\nc{\ee}{\eea}

\def\slash#1{\setbox0=\hbox{$#1$}#1\hskip-\wd0\hbox to\wd0{\hss\sl/\/\hss}}

\def\href#1#2{{#2}}

\rnc{\a}{\alpha}
\nc{\ab}{\bar{\a}}
\nc{\ap}{\a^{+}}
\nc{\abm}{\ab^{-}}
\rnc{\b}{\beta}
\nc{\bb}{\bar{\b}}
\nc{\bbp}{\bb_{\zb}^{+}}
\nc{\bm}{\b_{z}^{-}}
\nc{\oa}{\overline{\a}}
\nc{\ob}{\overline{\b}}
\rnc{\gg}{\gamma}
\rnc{\d}{\delta}
\nc{\f}{\phi}
\nc{\fb}{\bar{\phi}}
\nc{\vf}{\varphi}
\nc{\p}{\psi}

\rnc{\c}{\chi}
\nc{\la}{\lambda}
\nc{\m}{\mu}
\nc{\n}{\nu}
\rnc{\o}{\omega}
\nc{\Om}{\Omega}
\rnc{\t}{\theta}
\nc{\eps}{\epsilon}
\rnc{\S}{\Sigma}
\nc{\F}{\Phi}
\nc{\trac}[2]{{\textstyle\frac{#1}{#2}}}
\nc{\ex}[1]{\mbox{e}^{\,\textstyle#1}}
\nc{\mat}[4]{\left(\begin{array}{cc}#1&#2\\#3&#4\end{array}\right)}
\nc{\som}[9]{\left(\begin{array}{ccc}#1&#2&#3\\#4&#5&#6\\#7&#8&#9%
\end{array}\right)}
\nc{\tr}{\mathop{\mbox{tr}}\nolimits}
\nc{\ad}{\mathop{\mbox{ad}}\nolimits}
\nc{\Tr}{\mathop{\mbox{Tr}}\nolimits}
\nc{\Det}{\mathop{\mbox{Det}}\nolimits}
\nc{\rk}{\mathop{\mbox{rk}}\nolimits}
\nc{\ra}{\rightarrow}
\nc{\Ra}{\Rightarrow}
\nc{\LRa}{\Leftrightarrow}
\nc{\ot}{\otimes}
\rnc{\ss}{\subset}
\nc{\nul}{\noindent\underline}
\nc{\non}{\nonumber\\}
\nc{\subs}[1]{{\vspace*{0.5cm}}%
{\noindent\underline{#1}}{\addcontentsline{toc}{subsection}{#1}}%
{\vspace*{0.3cm}}}
\nc{\zb}{\bar{z}}
\nc{\pik}{\Pi_{\lk}}
\nc{\pip}{\Pi_{+}}
\nc{\pim}{\Pi_{-}}
\nc{\pih}{\Pi_{\lh}}
\nc{\jz}{J_{z}}
\nc{\jzh}{\jz^{\lh}}
\nc{\jzp}{\jz^{+}}
\nc{\jzm}{\jz^{-}}
\nc{\del}{\partial}
\nc{\dz}{\del_{z}}
\nc{\dzb}{\del_{\bar{z}}}
\nc{\az}{A_{z}}
\nc{\azb}{A_{\bar{z}}}
\nc{\g}{g^{-1}}
\nc{\dw}{\Delta_{W}}
\nc{\Ad}{{\mbox{Ad}}}
\nc{\ks}{Ka\-za\-ma-\-Su\-zu\-ki}
\nc{\KS}{\ks}
\nc{\ksm}{\ks\ model}
\rnc{\AA}{{\Bbb A}}
\nc{\BB}{{\Bbb B}}
\nc{\CC}{{\Bbb C}}
\nc{\PP}{{\Bbb P}}
\nc{\cpm}{\CC\PP(m)}
\nc{\cpn}{\CC\PP(n)}
\nc{\cp}[1]{\CC\PP(#1)}
\nc{\gmn}{G(m,m+n)}
\nc{\gmnk}{\gmn_{k}}
\nc{\cO}{{\cal O}}
\nc{\bcO}{\bar{\cO}}
\nc{\bO}{\bar{O}}
\nc{\oQ}{\overline{Q}}
\begin{document}
\global\parskip=4pt
\makeatother\begin{titlepage}
\begin{flushright}
\rightline{hep-th/9801083}
\end{flushright}
\vspace*{0.5in}
\begin{center}
{\LARGE\sc One Loop Effects In Various Dimensions And D-Branes}\\
\vskip .3in
\makeatletter
\begin{tabular}{cc}
{\sc Chaouki Boulahouache}\footnotemark
&
{\sc George Thompson}\footnotemark 
\\[.1in]
Syracuse University       & ICTP\\
Syracuse, New York        & P.O. Box 586\\
13244-1130                & 34014 Trieste\\
U.S.A                     & Italy\\
\end{tabular}
\end{center}
\addtocounter{footnote}{-1}%
\footnotetext{e-mail: chaouki@@suhep.phy.syr.edu}
\addtocounter{footnote}{1}%
\footnotetext{e-mail: thompson@@ictp.trieste.it}
\addtocounter{footnote}{1}%
\vskip .50in
\begin{abstract}
\noindent We calculate some one loop corrections to the effective
action of theories in $d$ dimensions that arise on the dimensional
reduction of a Weyl fermion in $D$ dimensions. The terms that we are
interested in are of a topological nature. Special attention is given
to the effective actions of the super Yang Mills theories that arise on
dimensional reduction of the $N=1$ theory in six dimensions or on the
dimensional reduction of the $N=1$ theory in ten dimensions. In the
latter case we suggest an interpretation of the quantum effect as a
coupling of the gauge field on the brane to a relative background gauge
field. 
\end{abstract}
\makeatother
\end{titlepage}
\begin{small}
\end{small}

\setcounter{footnote}{0}

\section{Introduction}

The low energy excitations of (d-1)-branes are governed by world volume
super Yang-Mills theories that are obtained on dimensional reduction to
$d$ dimensions. For $d \geq 5$ these theories are non-renormalizable
and only make sense up to some scale. From the string theory point of
view one needs to introduce $\a'$ corrections, precisely those terms that
were ignored in restricting ones attention to the super Yang-Mills
theory in the first place. One can ask, however, if there are field
theoretic quantities that can be calculated that do not depend on the
cutoff that one sets on the theory (alternatively that do not require
$\a'$ corrections)? Natural candidates are topological terms, of 
Chern-Simons type, that depend neither on the metric nor on the coupling. 
One knows that such terms are protected, perturbatively at least, without
the need to invoke any supersymmetric non-renormalization theorems.
They only pick up one loop corrections. Furthermore, these terms do
not depend on the cutoff (and so will not depend on $\a'$).

The terms that we will obtain are, then, of Chern-Simons type, that is
they are like secondary characteristic classes. Such terms arise in
non-supersymmetric theories, however, for concreteness, we will
presume in most of the text that we are given a supersymmetric system.
Suppose, then, that we start with
$N=1$ super Yang-Mills theory in $D$ dimensions, where $D=3$, $4$, $6$
or $10$. The resulting theory in $d$ dimensions, obtained on
reduction, will contain one vector $A$ and $D-d$ scalars $\f^{a}$. We
presume that the theory is in the Coulomb branch, that is, all the
massless fields are in the Cartan sub-algebra 
${\mathbf T}$ of the gauge group.

The
terms we are interested in will involve the normalized scalar
field $\hat{\f}= \f/\sqrt{|\f|^{2}}$ not $\f$, where $\f \in {\mathbf
T}$, and all contractions are
with respect to the $D$ dimensional epsilon symbol suitably
factorised. How does a term of the required form arise?
The presence of the epsilon symbols means that they can only come from a
fermion loop\footnote{Unless there are anti-symmetric tensors of
particular rank present}, for even $D$ with Weyl fermions one has
\be
\Tr \frac{(1+\Gamma_{D+1})}{2}.\Gamma_{\mu_{1}} \dots \Gamma_{\mu_{d}}
\Gamma_{a_{1}} \dots \Gamma_{a_{D-d}} \sim \eps_{\mu_{1} \dots
\mu_{d}}\eps_{a_{1} \dots a_{D-d}}.
\ee
To contribute to the effective action the fermions must be charged
with respect to the maximal torus (and hence massive). The origin of
these terms is much like that of Chern-Simons terms that appear in odd
dimensions associated with parity non-invariance \cite{red}.

With the conditions specified above what type of terms can we have? We
need to integrate a top form and so the number of times the gauge
field, in the Cartan sub-algebra, appears without a derivative (call it
$\a$), plus twice the number of factors of the field strength ($\gamma$)
together with the number of derivatives ($\b$) on $\f$ must equal the
dimension; $\a + 2\gamma + \b = d$. To soak up the internal
labels of the epsilon symbol there must be $(D-d)$ factors of
$\hat{\f}$. Because of the epsilon symbol there will be one
$\hat{\f}$ appearing without a derivative. Let $\sigma^{n}$ denote the
pull back of the unit volume form for the n-sphere $S^{n}$, 
\be
\sigma^{n} \sim \eps^{a_{1} \dots a_{n+1}}
d\hat{\f}_{a_{1}} \dots d\hat{\f}_{a_{n}} \, \hat{\f}_{a_{n+1}}, \;\;\; d
\sigma^{n} = 0.
\ee
The scalars will enter in the combination $\sigma^{D-d-1}$ so that $\b
= D-d-1$. Here $\sigma^{0} = {\mathrm sign}(\f)$. Likewise at most one
factor of the gauge field without a derivative can appear, i.e. $\a
=0$, $1$. Consequently, $\a + 2\gamma = 2d - D +1$ and if $D$ is even
$\a=1$, while if $D=3$ we have $\a =0$. Let us consider $D=3$ and $D=4$
separately. The terms of interest are potentially
\begin{center}
\begin{tabular}{||c|c|c||}\hline
d & D=3 & D=4 \\ \hline
3 &  & $\int A F \sigma^{0} $ \\ \hline
2 & $\int  F \sigma^{0}$ & $\int A \sigma^{1}$  \\ \hline
1 & $\int \sigma^{1} $ & \\ \hline
\end{tabular}
\end{center}
For $D=4$ we see that the putative terms are odd under the Weyl group
and so cannot arise from the vector multiplet. In section \ref{remarks} we
will present an argument that extends this conclusion to $D=4m$. For
$D=3$ the trace of 3 gamma matrices will give the sought for
epsilon symbols and those terms that arise from reductions of the
$D=3$ theory are Weyl even and so could, in
principle, appear. However, as can be seen explicitly, they do not
arise in the one loop calculation. Indeed, once more in section
\ref{remarks}, we will argue that for $D$ odd a massless Dirac fermion
coupled only to a gauge field cannot yield these topological terms
upon reduction.

For those theories that arise on the reduction from the $N=1$ SYM in
six and ten dimensions, one finds two and four derivatives terms
respectively:
\begin{equation}
\begin{tabular}{||c|c||}\hline
d & D=6 \\ \hline
5 & $\int A F F \sigma^{0} $ \\ \hline
4 & $\int A F\sigma^{1} $ \\ \hline
3 & $\int A \sigma^{2} $ \\ \hline
\end{tabular}\;\;\;\;\;\;\;\;\;
\begin{tabular}{||c|c||}\hline
d & D=10\\ \hline
9 & $\int A F^{4}  \sigma^{0} $ \\ \hline
8 & $\int A F^{3} \sigma^{1} $ \\ \hline
7 & $\int A F^{2} \sigma^{2} $ \\ \hline
6 & $\int A F\sigma^{3} $ \\ \hline
5 & $\int A \sigma^{4} $ \\ \hline
\end{tabular} \label{table}
\end{equation}

There is a pattern in these tables. For both types of reduction (from
$D=6$ or $D=10$) in the highest dimension ($d= 5$ or $d=9$) one
obtains a conventional Chern-Simons type term.  For each dimension one
descends a curvature form $F$ is replaced by the derivative of a
scalar, so that the degree of the volume form ascends, while keeping
the number of derivatives fixed. Also one can read from these tables
the analogues of the `peculiar' symmetry found by Seiberg \cite{seiberg5d},
namely there are conserved currents
\be
j = * \left( F^{d-D/2} \sigma^{D-d-1}    \right)
\ee
coupled to the gauge field itself.

So the `topological' terms that we are interested in involve only the epsilon
symbol and do not depend on the metric. In the supersymmetric field
theories under consideration such terms can only arise from fermion
loops and so we look for them in the fermion determinant contribution to the
effective action,
\be
\Tr \ln{ \slash{D}} = i \, \Gamma_{1-{\mathrm loop}}. \label{effact}
\ee
We perform a one loop computation, there are various
non-renormalization arguments, that will be recalled below, which
justify this at the perturbative level. Non-perturbative effects, can
and do contribute.

We begin by considering theories that arise on dimensional reduction
of the $N=1$ theory in $D=6$. These are perhaps the best studied
theories and much is known about them. Consequently for $d=4$ and
$d=5$ the results we find are already well known. The, minor, novelty
for $d=4$ is that the pre-potential is determined, perturbatively from
the one-loop correction to the $\theta$ angle rather than to the
$\beta$ function, these two ways of proceeding being equivalent once
one assumes the pre-potential is holomorphic. For $d=3$ it was
suggested in \cite{seiberg3d,SW3d} that a one loop calculation should
reproduce an index calculation for determining the metric on the
moduli space and here we see that this is true. 

We are rather more
brief about the results obtained from the reduction of the $N=1$
theory in $D=10$. The problem here is that the topological terms are
four derivative terms and so their direct geometric significance for
the moduli space is less clear. In terms of branes, the topological
terms correspond to an electric
coupling of background Ramond-Ramond fields to the gauge field. These
are not the Wess-Zumino terms, determined on consistency grounds for
T-duality, that are added to the D-brane action \cite{br,ght}. Rather,
though they they have a similar form, they are induced by the world
volume theory itself. 

We would have liked to give a more detailed
description of these topological objects in terms of branes (which is
where our motivation comes from) but, unfortunately, we have not
succeeded in finding a completely satisfactory description. Instead
we have to content ourselves with the above facts and that these are of an
interesting structure purely in field theory terms. On that note, we
show how these may arise starting with some Weyl fermion theory in
arbitrary dimension $D= 4m +2$.

One comment about the notation. All the theories under consideration
arise from the dimensional reduction of a theory in $D$
dimensions involving a Weyl fermion\footnote{When $D$ is odd we will
see that we do not find the terms we are looking for.}. Throughout we will
make use of the $D$ dimensional gamma
matrix algebra and, since the spinors are chiral, we
include a projection $(1 + \Gamma_{D+1})/2$ in gamma matrix traces. For
supersymmetric theories for the vector multiplets we insert $(1 +
\Gamma_{D+1})/2$ and a projection $(1 -
\Gamma_{D+1})/2$ for the hyper-multiplets. For the $N=1$ theory in
$D=10$ as the spinors are also Majorana we need to divide by a
further factor of $2$. 
Our conventions are that the Lie algebra valued fields appearing are
real, the generators $T$ are hermitian so that
\be
D_{M}=\partial_{M}-iA_{M},
\ee
where $A_{M}=A_{M}.T$, and $A_{M}^{\dagger}= A_{M}$. The metric in $D$
dimensions is $\eta_{MN} = \eta_{\mu \nu} \oplus -\d_{ab}$, however,
when the $D-d$ dimensional labels, $a\, , b \, , \, \dots$, appear
explicitly contraction will be with respect to $\d_{ab}$. This means
that $\Gamma_{M} = (\Gamma_{\mu}, \Gamma_{a})$ and $\Gamma^{M} =
(\Gamma^{\mu},  -\Gamma^{a})$, where $\Gamma^{a} = \d^{ab}\Gamma_{b}$.

\section{Reduction From Six Dimensions}

The $N=1$ theory in six dimensions is composed of a vector $A_{M}$ and
a Weyl spinor $\Psi$. This theory has an $SU(2)_{{\cal R}}$
$R$-symmetry group that acts only on the fermion. When we reduce to
$d$ dimensions that Lorentz group decomposes as $SO(5,1) \supset
SO(d-1,1) \otimes SO(6-d)$. The vector decomposes as $A_{M} = A_{\mu}
\oplus \f_{a}$, with $\mu = 0, \dots, d-1$ and $a= 1, \dots 6-d$. The
$R$ symmetry is enhanced to $SO(6-d)\otimes SO(3)_{{\cal R}}$ but
which now also acts on the scalars $\f$. The theories that we are
considering here have an interpretation as configurations of various
intersecting branes.

\subsection{The Calculation}

The variation of the one loop effective action, with $N_{f}=0$, with
respect to the gauge field is\footnote{The trace here can extend
over the entire Lie algebra as the Cartan components will not
contribute to (\ref{var}) in any case.}
\be
i \d \Gamma_{{\mathrm 1-loop}} = -i \int d^{d}x \, \Tr \, \ad \left(\d
\slash{A} \right) <x|
\frac{1}{ \Gamma^{M}D_{M}} |x> . \label{var}
\ee
We need to make use of the Lichnerowicz formula for the square of the
Dirac operator in six dimensions,
\bea
\slash{D}^{2} &=& D^{M}D_{M} - \frac{i}{2} \Gamma^{M}\Gamma^{N} \ad
\left( F_{MN} \right) \nonumber \\
&=& D^{M}D_{M} - \frac{i}{2} \Gamma^{\mu}\Gamma^{\nu} \ad
\left( F_{\mu \nu} \right) + i \Gamma^{\mu}\Gamma^{a}\partial_{\mu}
\ad \left( \f_{a} \right) ,
\eea
where
\be
D^{M}D_{M} = D^{\mu}D_{\mu} + \ad \left( \f_{a} \right) \ad \left(
\f^{a} \right) . \label{Dsq}
\ee
Since every field appearing is taken to live in the Cartan sub-algebra
the commutators $[\f , \f]$ and $[A_{\mu}, \f ]$ are zero.
We wish to pick out of
\be
i \d \Gamma_{{\mathrm 1-loop}} = -i \int d^{d}x \, \Tr \,  \ad
\left(\d \slash{A}\right) . <x|
\frac{\slash{D}}{ \slash{D}^{2} } |x> , \label{var2}
\ee 
the term proportional to the epsilon symbol and not involving the metric.
It should be clear from (\ref{Dsq}) that, in this context, we can
safely replace $D^{\mu}D_{\mu}$ with $\partial^{\mu}\partial_{\mu}$.
To obtain an epsilon symbol we should exactly saturate the number of
$\Gamma$ matrices that appear in $\Gamma_{7}$. From the numerator
$\slash{A}$ cannot contribute since this would mean that the gauge
field would appear twice without derivatives and such a term would
vanish when contracted with the epsilon symbol. Likewise, the
$\slash{\partial}$ term in the numerator cannot contribute since, by
an integration by parts, this would lead to a contribution where the
gauge field always appears in the form of a field strength and a glance
at the first table of (\ref{table}) shows us that we are in search of
terms that involve the gauge field without derivatives. Consequently,
we may as well focus on
\be
i \d \Gamma_{{\mathrm 1-loop}} =  \int d^{d}x \, \Tr \,  \ad\left(\d
\slash{A}\right)  . \ad
\left( \Gamma^{a}\f_{a} \right) . <x|
\frac{1}{ \slash{D}^{2} } |x> + \dots \; . \label{var3}
\ee

One must now expand the denominator so as to pick up a term with the
field strength raised to the $d-3$ and $\partial_{\mu}\f_{a}$ raised
to the power of $5-d$. A further simplification is that one does not
need to worry about operator ordering since any error introduced will
be of a higher order in derivatives or/and will involve the metric.
Hence, 
\bea
\frac{1}{ \slash{D}^{2} } &=&
\sum_{n=0}^{\infty}\frac{i^{n}\left(\frac{1}{2}\Gamma^{\mu}\Gamma^{\nu}
\ad \left(F_{\mu  \nu}\right) -  \Gamma^{\a} \Gamma^{a} \partial_{\a}
\ad \left(\f_{a}\right)
\right)^{n}}{ \left( \partial^{2}  + \ad \left( \f \right)^{2}
\right)^{n+1}} \nonumber \\
& =& \dots \nonumber\\
&& +  \frac{(-1)^{d}2!}{2^{d-3}(5-d)!(d-3)!}
\frac{\left(\Gamma^{\mu}\Gamma^{\nu} \ad \left(F_{\mu
\nu} \right) \right)^{d-3} \left(
\Gamma^{\a} \Gamma^{a} \partial_{\a} \ad \left(\f_{a}\right) 
\right)^{5-d}}{\left(\partial_{\f }^{2} \right)^{3}} \nonumber\\
&&+ \dots , \label{expand6}
\eea
where
\be
\partial_{\f}^{2} = \partial^{2} + \left(\ad (\f) \right)^{2} .
\ee
The error in passing the $F$ terms through the $d\f$ involves
anti-commutators of gamma matrices and so metric pieces which we can
safely ignore. The error in passing these operators through the Greens
functions $\partial_{\f}^{-2}$ involves higher order derivatives
which, to the order that we are working, can also be ignored. 

We see from (\ref{expand6}) that as $d$ increases we exchange the
field strength $F_{\mu \nu}$ for the derivative of the scalar
$\slash{\partial}\f$. What is being held fixed is, from the six
dimensional point of view, the term in the
Taylor series proportional to $ (\Gamma^{M}\Gamma^{N}F_{MN})^{2}$.

The only
integral that needs to be evaluated is
\bea
<x| \frac{1}{\left(\partial_{\f}^{2}\right)^{3}} |x> &=& - \int
\frac{d^{d}p}{ \left( 2 \pi \right)^{d}}\; \frac{1}{\left( p^{2}- \left(\ad
( \f)\right)^{2} \right)^{3}} \nonumber \\
& =& \frac{i}{2}
\frac{\Gamma(3-\frac{d}{2})}{(2\sqrt{\pi})^{d}}\frac{1}{|\ad
(\f)|^{6-d}} . 
\eea
One may put the pieces together to determine the contribution to the
effective action. We do this for for the individual cases in the
following subsections. In order to obtain the $1$-loop effective action
contribution of the
hyper-multiplets one simply exchanges the $\ad$ representation in the
formulae above with the appropriate one and multiplies by the
ubiquitous $-1$. 

\subsection{$N=4$ SYM In $d=3$}

Seiberg began the study of the Coulomb branch of the $N=4$ theories in
three dimensions \cite{seiberg3d}. He suggested, that for $U(1)$ and
$SU(2)$ gauge groups that it would be possible to determine the metric
on the moduli space by performing a one loop computation. This was
re-iterated in a work by Seiberg and Witten \cite{SW3d}, however,
these authors determined the metric on the moduli space, for the two
groups above, by counting zero modes in a monopole field. They found
that the metric corresponds to that of a hyper K\"{a}hler manifold
which, at infinity, is (a quotient of) the Lens space ${\mathbf L}_{-4}$.
This space, for $SU(2)$, is the classical moduli space of BPS
monopoles in the $SU(2)$ theory.

For the metric on this space to be hyper-K\"{a}hler it is necessary
that the coupling constant, and hence the metric on ${\mathbf R}^{3}$ is
corrected to 
\be
\frac{1}{e^{2}} \to \frac{1}{e^{2}} - \frac{s}{|x|} .
\ee
That this is indeed the case has been established explicitly in
\cite{dkmtv1}.

The Coulomb branch of the $SU(n)$ theories was investigated by
Chalmers and Hanany \cite{CH3d} and they suggested that the moduli
space corresponds to the centered moduli space of n monopoles. Once
more the metric was determined by the use of the Callias index formula
\cite{Callias}. The index turns out not to depend on $n$ and one finds
once more that $s_{3} =-4$. This is consistent with the perturbative
formula that we have obtained above as we will now see.

The term, in the effective action, that we are after is\footnote{The
normalisation here is chosen so that $s_{3}^{\mathrm adj}=1$ corresponds
to the degree one monopole bundle over the two sphere.}
\bea
i \Gamma_{{\mathrm 1-loop}} &=& -\sum_{{\mathbf R}} \frac{is_{3}^{{\mathbf
 R}}}{32\pi}\;
\eps^{\mu \nu \lambda}\eps^{abc} \, \Tr_{{\mathbf R}}
\int d^{3}x \; A_{\mu} \, \partial_{\nu}\hat{\phi}_{a}\, 
\partial_{\lambda}\hat{\phi}_{b} \, \hat{\phi}_{c} \nonumber \\
& = & \sum_{{\mathbf R}}\, s_{3}^{{\mathbf R}}I_{{\mathbf R}} , 
\label{1loop3d}
\ee
where ${\mathbf R}$ is the representation of the fermions. The
perturbative calculation yields,
\bea
&&i\d \Gamma_{{\mathrm 1-loop}}^{{\mathrm adj}} = \nonumber\\
&&\frac{i}{8\pi}\eps^{\la \mu
\nu }\eps^{abc} \Tr \, \int d^{3}x
\; \ad (\d A_{\la}) \ad\left(\hat{\f}_{c}\right)\,
\partial_{\mu}\ad\left( \hat{\f}_{a} \right) \,
\partial_{\nu}\ad \left(\hat{\f}_{b}\right) \, + \dots \; ,
\eea
or
\bea
&&i\Gamma_{{\mathrm 1-loop}}^{{\mathrm adj}} = \nonumber\\
&&\frac{i}{8\pi}\eps^{\la \mu \nu
} \eps^{abc} \Tr \, \int d^{3}x
\; \ad (A_{\la}) \ad\left(\hat{\f}_{c}\right)
\partial_{\mu}\ad\left(\hat{\f}_{a}\right)
\partial_{\nu}\ad \left(\hat{\f}_{b}\right) \, + \dots \; .
\eea

The theory with no
hyper-multiplets has for its only non-zero contribution $s_{3}^{{\mathrm
adj}}=-4$. The hyper-multiplets all contribute $s_{3}^{{\mathbf R}} =
4$. For $SU(2)$ and a 
theory with $N_{f}$ hyper-multiplets in the fundamental representation
we find that there is a universal factor $s_{3}=N_{f}s_{3}^{{\mathrm
fund}}/2 +s_{3}^{{\mathrm adj}} =
2N_{f}-4$ times the adjoint space factor $I_{{\mathrm adj}}$. This is in
agreement with the index calculation for the number of zero modes in
the presence of monopoles. The relative factor of $2$ can be
understood from the following considerations. One can scale $\f$
freely in (\ref{1loop3d}) so the only thing we need to declare is the
charge pre-factor of the gauge field. The charge in the adjoint
representation is twice that in the fundamental. Looking forward a bit
we see that this argument tells us that in four dimensions (where the
gauge field appears twice) one obtains a factor of $(8-2N_{f})$ and in
five dimensions it is $(16 - 2N_{f})$.

Apart from non-perturbative effects it is argued in \cite{SW3d} that
one loop is exact. So we need look only at the term in the effective
action that arises on integrating out the massive (and charged)
fermions. For the theory with $N_{f}=0$, this means that we are
interested in (\ref{effact}) with $\slash{D}$ acting on adjoint
valued fermions.

Let us be explicit about the formula for the vector multiplet. Let
$\a \in \Delta^{+}$ be the positive roots of the algebra. We have
\bea
& & \frac{i}{8\pi}\;
\eps^{\mu \nu \lambda}\eps^{abc} \, \Tr
\int d^{3}x \; \ad (A_{\mu}) \, \partial_{\nu}\ad (\hat{\phi}_{a})\, 
\partial_{\lambda}\ad (\hat{\phi}_{b}) \, \ad (\hat{\phi}_{c}) \nonumber
\\ 
& & \;\;\; =
\frac{i}{4\pi} \eps^{\mu \nu \lambda}\eps^{abc} \,
\sum_{\a \in
\Delta^{+}} \int d^{3}x \, \a (A_{\mu}) \, \partial_{\nu}\a (
\hat{\phi}_{a}) \,
\partial_{\lambda}\a( \hat{\phi}_{b}) \, \a( \hat{\phi}_{c}) ,
\eea
with
\be
\a(\hat{\f}_{a}) = \frac{\a(\f_{a})}{\sqrt{\a(\f_{b})\a(\f^{b})}} .
\ee
The fields, $\a(\hat{\f}_{i})$, that appear define $S^{2}$'s in the
moduli space. Notice that 
\bea
\sigma_{\a} &=& \frac{1}{8\pi} \eps^{abc} \, d \a (\hat{\f}_{a})\, d \a
(\hat{\f}_{b}) \, \a (\hat{\f}_{c}) \nonumber \\
& = & \left(\sigma_{\a }\right)_{ \mu \nu}\; dx^{\mu} dx^{\nu}
\eea 
is the unit volume form for the two sphere defined by $\a
(\hat{\f}_{a})$.

There are exactly a positive number of roots of
such spheres\footnote{Since the negative roots just double the
contribution of the positive roots we do not need to consider them
independently.}, i.e. there are
\be
\frac{1}{2} ({\mathrm dim}{\mathbf G}- {\mathrm rank}{\mathbf G})
\ee
such spheres. For $SU(n)$ this is $n(n-1)/2$. These will play a role
subsequently.

In three dimensions each vector is dual to a compact scalar. The
moduli space is classically, before quotienting with the Weyl group,
$\left( {\mathbf R}^{3} \times S^{1}\right)^{r}$. This is a $4r$
dimensional manifold with an $SO(3)$ action on it. It is
naturally a hyper-K\"{a}hler manifold. The one-loop result that we
have just calculated shows that the moduli space no longer has a
product structure. Rather we saw that a number of spheres are
appearing and the end result is that at ``infinity'' one finds not
spheres times circles, but rather circle bundles over the spheres. To
see this we sketch how the dualisation is to be performed.

One considers the effective action up to one loop order as a function
of a ranks worth of arbitrary two forms $F^{i}$ and one adds Lagrange
multipliers $\theta_{i}$ imposing the constraints that the $F^{i}$ are
closed\footnote{The $\theta_{i}$ are also taken to be periodic, so
that the harmonic parts of $F^{i}$ are integral.}.
Explicitly the action is, 
\be
\int d^{3}x \; g_{ij}\left( - \frac{1}{4}F^{i}_{\mu \nu} F^{j \, \mu
\nu} + \frac{1}{2} \partial_{\mu}\f^{i} \partial^{\mu} \f^{j} \right)
+\frac{i}{8\pi} \eps^{\mu \nu \lambda } F_{\mu \nu}^{i}\left(
\partial_{\lambda} \theta_{i} -  \beta_{ij}^{a}
\partial_{ \lambda}
\hat{\f}^{j}_{a} \right) . \label{dual}
\ee
If one integrates out $\theta_{i}$ one finds the required constraint
$dF^{i}=0$ whose solution is taken to be $F^{i}= dA^{i}$. Once one
substitutes this back into the action (\ref{dual}) one re-obtains the
one loop corrected effective action. The $g_{ij}$ represent the
coupling constants up to one loop, which we have not calculated, see
\cite{dkmtv1}, while the term involving
$\beta_{ij}^{a}$ is, by definition,
\be
\int d^{3}x \, i \eps^{\mu \nu \lambda} F_{\mu \nu}^{i}\beta_{ij}^{a}
\partial_{\lambda} \hat{\f}^{j}_{a} = -\sum_{{\mathbf R}} \,
is_{3}^{{\mathbf R}} \;
\eps^{\mu \nu \rho}\eps^{abc} \, \Tr_{{\mathbf R}}
\int d^{3}x \;  A_{\mu} \, \partial_{\nu} \hat{\phi}_{a}\, 
\partial_{\rho} \hat{\phi}_{b} \,  \hat{\phi}_{c} .
\ee

Now, integrating out the fields $F^{i}$ give rise to a dual action,
from which we obtain the metric. This is
\be
\int d^{3}x \, \frac{1}{2} g_{ij}\partial_{\mu}\f^{i} \partial^{\mu}
\f^{j} +2 g^{ij}\left(
\partial_{\lambda} \theta_{i} - \beta_{ik}^{a} \partial_{\lambda}
\hat{\f}^{k}_{a} \right)\left(
\partial^{\lambda} \theta_{j} - \beta_{jl}^{b} \partial^{\lambda}
\hat{\f}^{l}_{b} \right) , \label{dual2}
\ee
where $g^{ij}$ is proportional to the inverse matrix of $g_{ij}$. From
(\ref{dual2}) we see that the product structure with the circles has
been turned into a monopole bundle. 

While the formulae above are quite general we would now like to make
a brief comparison with the metric of \cite{CH3d}. These authors
determined the metric for the $SU(n)$ moduli space. If one denotes the
moduli space by ${\cal M}_{SU(n)}$, they show that $H_{2}({\cal
M}_{SU(n)} , {\mathbf Z}) = n(n-1)/2$ and, furthermore, they determined
that $s_{3}^{{\mathrm adj}}=-4$ by computing the Callias index in the
presence of a monopole. The spheres that appear are the same two spheres
that we see in the one-loop expression. Furthermore, the one loop
computation also tells us that $s_{3}^{{\mathrm adj}}=-4$. 
Put another way; one does not need to feed in information about the
moduli space, rather the one-loop correction ``knows'' about the
structure of ${\cal M}_{SU(n)}$.

\subsection{$N=2$ SYM In $d=4$}

The one loop contribution to the topological terms is
\bea
&&i \Gamma_{{\mathrm 1-loop}} = \nonumber\\
&&\sum_{{\mathbf R}}\frac{s_{4}^{{\mathbf
R}}}{64\pi^{2}}  \eps^{\mu
 \nu \la \rho}\eps^{ab} \, \Tr_{{\mathbf
R}} \int d^{4}x \; \ad (A_{\mu}) \, \partial_{\nu} \ad (A_{\la}) \,
\partial_{\rho}\ad (\hat{\phi}_{a}) \, \ad (\hat{\phi}_{b}) . \label{1loop4d}
\eea
{}From the perturbative calculation we have found that
\be
i \d \Gamma_{{\mathrm 1-loop}}^{{\mathrm adj}} = 
\frac{i}{8\pi^{2}} \eps^{\mu \nu
\rho \la} \eps^{ab} \, \Tr \, \int d^{4}x \, \ad (\d A_{\mu}) \ad
(F_{\nu \rho}) \ad(\partial_{\la}\hat{\f}_{a}) \ad(\hat{\f}_{b}) +
\dots \, , 
\ee
or
\be
i \Gamma_{{\mathrm 1-loop}}^{{\mathrm adj}} = 
\frac{i}{16\pi^{2}} \eps^{\mu \nu
\rho \la} \eps^{ab} \, \Tr \, \int d^{4}x \, \ad ( A_{\mu}) \ad
(F_{\nu \rho}) \ad(\partial_{\la}\hat{\f}_{a}) \ad(\hat{\f}_{b}) +
\dots \; . \label{4dact}
\ee
{}From this we see that $s_{4}^{{\mathrm adj}}=8$ and the
hyper-multiplets give $s_{4}^{{\mathbf R}}=-8$. 

Let us fix on $SU(2)$ and write $\hat{\f}_{a} = (\cos{\varphi},
\sin{\varphi})$ then we have
\be
i \Gamma_{{\mathrm 0 + 1-loop}}^{{\mathrm adj}} = 
\frac{i}{32\pi} \eps^{\mu \nu
 \la \rho}  \,  \int d^{4}x \,
\left(\frac{\theta}{2\pi}-\frac{2\varphi}{\pi}  \right)
F_{ \mu \nu} F_{\la \rho}  +
\dots \; . \label{theta}
\ee
This agrees with the exact perturbative result found long ago
\cite{sv1,sv2,sv3}. To see this we recall that the one loop correction
to the pre-potential, in the notation of \cite{sw4d}, is
\be
{\cal F}_{{\mathrm 1-loop}} =
\frac{i}{2\pi}a^{2}\ln{\frac{a^{2}}{\Lambda^{2}}}.
\ee
Consequently the one loop correction to $\theta / 2\pi$ is
\be
{\mathrm Re} 
\frac{\partial^{2}{\cal F}_{{\mathrm 1-loop}}}{\partial a^{2}} =
{\mathrm Re} \frac{2i}{\pi} \ln{\frac{a}{\Lambda}} ,
\ee
and, as $a \sim \ex{i\varphi}$, one sees that this agrees with
(\ref{theta}). While here we have only determined the one loop
correction to the theta term, using super-symmetry and holomorphy one
may deduce the complete form of the one-loop corrected pre-potential.

For a general group we may re-write (\ref{4dact}) as
\be
i\Gamma_{{\mathrm 1-loop}} = - \frac{i}{16\pi^{2}} \eps^{\mu \nu \rho
\la} \sum_{\a \in \Delta^{+}} \int d^{4}x \, \a \left( F_{\mu \nu}
\right) \, \a \left( F_{\rho \la} \right) \varphi_{\a}
\ee
where $\a(\hat{\f}_{a})= (\cos{\varphi_{\a}}, \sin{\varphi_{\a}})$.
{}From this we may deduce that
\be
{\mathrm Re}\, \frac{\partial^{2}{\cal F}_{{\mathrm 1-loop}}}{\partial a^{i}
\partial a^{j}} = - \frac{2}{\pi} \sum_{\a \in \Delta^{+}} \a_{i}
\a_{j} \varphi_{\a} , 
\ee
or that
\be
{\cal F}_{{\mathrm 1-loop}} = \frac{i}{2\pi} \sum_{\a \in \Delta^{+}}
\a(a)^{2} \ln{\frac{\a(a)^{2}}{\Lambda^{2}}} .
\ee
Once more we have not at all controlled non-perturbative effects. 

\subsection{SYM in $d=5$}
Seiberg \cite{seiberg5d} initiated a study of non-renormalizable field
theories in five (and six) dimensions (see also \cite{more5d,more5d2}). As
they are non-renormalizable these theories must be understood to come
with some regularization, say a cutoff or dimensional regularization.
The term that we are interested in is, in any case, finite so we do
not really have to specify which except to say that it is a gauge
invariant regularization. The ``topological'' term that can be
generated in five dimensions is of the form
\bea
&&i \Gamma_{{\mathrm 1-loop}}= \nonumber\\
&&-\sum_{{\mathbf R}}\frac{is_{5}^{{\mathbf R}}}{48
\pi^{2}} \eps^{\mu \nu \la
\rho \sigma} \, \Tr_{{\mathbf R}} \int d^{5}x \;
\ad (A_{\mu}) \ad (\partial_{\nu} A_{\la}) \, \ad
(\partial_{\rho}A_{\sigma}) {\mathrm sign}(\ad (\phi)) \nonumber\\
&&+ \dots \; . \label{1loop5d}
\eea

In the case of $U(1)$, Witten \cite{witten5d} has performed the
calculation. The result is $s_{5}(U(1)) = -N_{f}$. For a non-Abelian
group ${\mathbf G}$ with $N_{f}=0$ we have
\be
s_{5}^{{\mathrm adj}}({\mathbf G}) =1 .
\ee

Some remarks are in order. As
explained in \cite{witten5d} the sign of the mass term in the $U(1)$
theory is observable. Hence, different signs correspond to different
theories. On the other-hand, in the non-Abelian theory (without $U(1)$
factors) that sign is {\bf not} observable. Rather, ${\mathrm sign}(\ad
(\phi))$ appears in (\ref{1loop5d}) as a consequence of gauge
invariance. After restricting to the Cartan sub-algebra the Weyl group
acts by permutation and (\ref{1loop5d}) is indeed Weyl invariant. As
an example, for $SU(2)$ the Weyl group acts by sending the Cartan
generator to minus itself so that $A_{\mu} \to - A_{\mu}$ and
$\Gamma_{{\mathrm 1-loop}}$ is invariant since ${\mathrm sign}(\ad (\phi))$
also changes sign. Now since the theory is Weyl invariant we can focus
on the moduli space ${\mathbf R}^{{\mathrm rank}}/{\mathrm Weyl}$, 
which in the
case of $SU(2)$ is ${\mathbf R}^{+}$.

\section{Reduction From Ten Dimensions}

The $N=1$ theory in ten dimensions is made up of a vector $A_{M}$ and
a Majorana-Weyl spinor $\Psi$ and consequently there is no $R$
symmetry here. On reduction to $d$ dimensions the vector decomposes as
$A_{M} = A_{\mu} \oplus \f_{a}$ where $\mu = 0, \dots, d-1$ and $a=1
\dots 10-d$. The Lorentz group decomposes as $SO(9,1) \supset
SO(d-1,1) \otimes SO(10-d)$ and $SO(10-d)$ is the $R$ symmetry group
acting both on the spinor and on the scalars. 

The interpretation of these theories in terms of branes is quite
direct. The $U(n)$ theory in $d$ dimensions arises as $n$ $d-1$ branes
come together. The broken theory, where the scalars have non-zero
expectation values, correspond to the branes being separated. If one
ignores the center of mass motion the theory of $n$ separated
branes has a world volume $SU(n)$ gauge symmetry. Of course in what
follows any gauge group may be chosen.

\subsection{The Calculation in $D$ dimensions}
We will perform the calculation in a little more generality than is
required in this section as it will prove useful later. The dimension
$D$ is taken to be even but otherwise arbitrary. Nevertheless,
the calculation is almost identical to the one we performed before so
we can be brief about it. As before
\be
i\d \Gamma_{{\mathrm 1-loop}}^{{\mathrm adj}} = 
\Tr \int d^{d}x \ad(\d \slash{A})
\ad( \Gamma^{a}\f_{a}) <x| \frac{1}{\slash{D}^{2}} |x> ,
\ee
and one expands $\slash{D}^{-2}$ out to the following order, once
again ignoring questions of ordering,
\bea
\frac{1}{\slash{D}^{2}} &=& \dots 
+ \frac{\left(i^{D/2-1} \frac{1}{2}\Gamma^{\mu}
\Gamma^{ \nu}F_{\mu \nu} - \Gamma^{\mu} \Gamma^{a}\ad(\partial_{\mu}
\f_{a})\right)^{D/2-1}}{\left( \partial_{\f}^{2} \right)^{D/2}} + \dots
\nonumber \\
&=& \dots + \left( \begin{array}{c}
D/2-1 \\
d-D/2
\end{array}\right)
 \frac{(\frac{1}{2}\Gamma^{\mu}
\Gamma^{ \nu}F_{\mu \nu} )^{d-D/2} (- \Gamma^{\mu}
\Gamma^{a}\ad(\partial_{\mu} \f_{a}) )^{D-d-1}}{\left( \partial_{\f}^{2}
\right)^{D/2}} + \dots \; .
\eea 

The integral that needs to be performed is
\bea
<x| \frac{1}{\left( \partial_{\f}^{2} \right)^{D/2}} |x> &=& 
(-1)^{D/2} \int
\frac{d^{d}p}{(2\pi)^{d}} \frac{1}{\left( p^{2}- (\ad (\f))^{2}
\right)^{D/2}} \nonumber \\
&=& \frac{i}{\Gamma(D/2)}.\frac{\Gamma(\frac{D}{2}-\frac{d}{2})}{(
2\sqrt{\pi} )^{d}}.\frac{1}{|\ad
(\f)|^{D-d}}.
\eea
Chirality is taken to be defined with respect to the projector $(1 +
\Gamma_{D+1})/2$ where,
\be
\Gamma_{D+1} = (-i)^{(D/2-1)}\Gamma_{0}\Gamma_{1}\dots \Gamma_{D-1} .
\ee
It is convenient also to let $d=2k$ if $d$ is even and $d=2k+1$
if odd.

For a Weyl fermion in $D$ dimensions the one loop effective action in
$d$ dimensions is
\be
\Gamma_{{\mathrm 1-loop}}^{{\mathrm adj}}(D,d)  = iC_{(D,d)} \Tr \int
d^{d}x \, G(A)^{\rho_{1} \dots \rho_{D-d-1}} \, 
\sigma(\hat{\f})_{\rho_{1} \dots \rho_{D-d-1}} ,\label{Deffective}
\ee
where
\bea
G(A)^{\rho_{1} \dots \rho_{D-d-1}}& = &\eps^{\la \, \mu_{1}\nu_{1}
\dots \mu_{d-D/2}\nu_{d-D/2} \,
\rho_{1}  \dots
\rho_{D-d-1}} . \nonumber \\
& & \;\; \, \ad (A_{\la})\, \ad
(F_{\mu_{1}\nu_{1}}) \dots \, \ad (F_{\mu_{d-D/2}\nu_{d-D/2}}),
\eea
\be
\sigma(\hat{\f})_{\rho_{1} \dots \rho_{D-d-1}}  =  \eps^{a_{1} \dots
a_{D-d-1} \, a} \, \partial_{\rho_{1}}
\ad(\hat{\f}^{a_{1}}) \dots \, \partial_{\rho_{D-d-1}}
\ad(\hat{\f}^{a_{D-d-1}}) \, \ad(\hat{\f}^{a}) ,
\ee
and the co-efficient $C_{(D,d)}$ is
\be
C_{(D,d)} = \frac{(-1)^{k-1}2^{D-d-1}}{(d+1-D/2)} \left(\begin{array}{c} 
D/2 -1 \\
d- D/2
\end{array}
\right)\frac{\Gamma(D/2-d/2)}{(2\sqrt{\pi})^{d}\Gamma(D/2)} \label{DC} .
\ee
With $D=6$ the results of the previous section are reproduced.

\subsection{Interpretation}

Now we set $D=10$ to arrive at the one loop effective action in $d$
dimensions 
\be
\Gamma_{{\mathrm 1-loop}}^{{\mathrm adj}}  = iC_{d} \Tr \int
d^{d}x \, G(A)^{\rho_{1} \dots \rho_{9-d}} \, 
\sigma(\hat{\f})_{\rho_{1} \dots \rho_{9-d}}
\eea
where the co-efficient $C_{d}$ is
\be
C_{d} = \frac{1}{2}C_{(10,d)}
\ee
since the $N=1$ fermions in $10$ dimensions are also Majorana. For
$SU(2)$, as we discussed before, the relative contribution of the
adjoint representation to the fundamental goes like
$(2^{d-3}-2N_{f})$. 

In order to give a possible interpretation of the result we recall some
facts about $p$-branes. To a single $p$-brane (of $D$ dimensional
super-gravity) one can
associate an electric charge, $Q_{E}$,
with respect to a background gauge field, $C^{p+1}_{D}$. For a
$p$-brane one has
\be
Q_{E} = \int_{S^{D-p-2}} * d C^{p+1}_{D} .
\ee
At large transverse distance, $r$, to the brane the form behaves as
\be
C^{p+1}_{D} \sim Q_{E}r^{p+3-D}\o_{p+1} ,
\ee
where $\o_{p+1}$ is the volume form on the brane. Hence,
\be
*dC^{p+1}_{D} = Q_{E} \sigma^{D-p-2}.
\ee

Now consider the situation of two well separated $(d-1)$ branes. We
recall that the expectation values of the scalar fields, on
dimensional reduction, measure the separation of the branes. They represent,
therefore, directions normal to the branes. Consequently the unit
spheres, with volume forms $\sigma^{p}$, are orthogonal to the branes.
Up to normalization, the one loop effective action has the form
\bea
\Gamma_{{\mathrm 1-loop}} &\sim & \int A F^{d-D/2}\sigma^{D-d-1}
\nonumber \\
& \sim & \int A F^{d-D/2} * d C^{d}_{D} \nonumber \\
& \sim & \int F^{d+1-D/2} \tilde{C}^{D-2 -d}_{D} ,
\eea
where $\tilde{C}^{D-2-d}_{D}$ is the magnetic dual of $C^{d}_{D}$, $*
d C^{d}_{D} = d \tilde{C}^{D-2 -d}_{D}$, and the volume forms are now
understood to be the pull backs to the brane. So we have found a
coupling between the gauge field on the brane and a `relative'
Ramond-Ramond background gauge field
$\tilde{C}^{D-2-d}(\hat{\f})_{D}$. When there are $n$ well separated branes,
the one loop effective action will involve couplings between the $n-1$,
$U(1)$ gauge fields and the $n(n-1)/2$ relative Ramond-Ramond
background gauge fields $\tilde{C}^{D-2-d}_{D}$. 

While we have not proven that this picture comes out of string theory,
we note that the numerology at least comes out right. The type IIA
$4$-brane carries a magnetic charge under the three-form $C$, so for
separated $4$-branes we can expect a relative three form background
field. The type IIA $6$-brane is, likewise, magnetically charged under the
type IIA one-form potential. The type IIB $5$-brane has a magnetic
charge under the Ramond-Ramond two form.

Notice that we may also dualise the five dimensional `photon'.
Ignoring all other quantum effects one obtains a metric on the
moduli space, at infinity, of the form
\be
(d\theta - \beta)^{2} ,
\ee
where $\theta$ is a compact
two-form, dual to the photon, and $d\beta = \sigma_{4}$.

\section{Concluding Remarks}\label{remarks}

Throughout the text we have concentrated on the reduction of
supersymmetric gauge theories. However, one may also consider
non-supersymmetric theories even though these may not make sense as
full fledged field theories for $d>4$. The conditions that we saw in the
introduction still need to be met, namely that
\bea
\a + 2\gamma + \beta & = & d \nonumber \\
\beta & =& D-d-1 , \label{conds}
\eea
but we must augment these with the condition that the action be Weyl
group even (indeed more generally Weyl invariant). The latter implies
that we always have an even number of fields appearing in the
effective action, 
\be
\a + \gamma + \beta + 1 = 2m
\ee
for some $m$. Let us take $D$ to be even, which implies that $\a =1$.
Then manipulation of (\ref{conds}) yields $D=4m + 2$. So up to $10$
dimensions the only one we overlooked in the text was $D=2$ with $d=1$. The
generic formulae for Weyl fermions were given in the preceding
section (\ref{Deffective}). What of $D$ odd? In such cases the
fermions are not Weyl,
however, the trace of $D$ gamma matrices will give the sought for
epsilon symbols. But, from the way the calculation proceeds, as in
(\ref{var2}), the trace is always over an even number of gamma
matrices and so the topological terms never arise. 

With a small variation on the theme one may recover the conventional
Chern-Simons actions. For example, in $d=3$ ($D=4$) such a term was
precluded by Weyl reflections. But suppose that the scalar field is
not in the adjoint of the Lie algebra but rather is proportional to
the identity. In this case $\f$ would be a mass term, $\sigma^{0}$
would not be Weyl odd and the Chern-Simons action would survive the
Weyl reflection. The co-efficient that one gets, $C_{(4,3)}=i/16\pi$
agrees with that found for the $U(1)$ theory in \cite{red}. A similar
story holds for other $d=4m-1$ ($D=4m$).

The reductions that we have been considering have all been
spatial. To obtain Euclidean field theories it makes sense to
perform a reduction in the time direction \cite{bt}. Such
reductions are especially useful in understanding how topological
field theories arise \cite{bsv,bt2,bbm,bks,bt,bbjm}. The fundamental 
difference with the spatial reduction is that now the scalars, $\f^{a}$, 
transform as a vector of $SO(D-d-1,1)$. The normalised scalars
$\hat{\f}$ do not define the sphere, $S^{D-d-1} \subset {\mathbf
R}^{D-d}$, but rather the hyperbloid, $H^{D-d-1} \subset {\mathbf
R}^{(D-d-1,1)}$. 

More generally, one may also ask what happens when one
wraps D-branes around non-trivial cycles of some Calabi-Yau manifold?
This has been answered in \cite{bsv}; some of the scalars become
sections of the normal bundle to the cycle in the Calabi-Yau manifold.
For us, this means that when computing one loop corrections we should
further decompose the $10$ dimensional epsilon symbol (corresponding
to the decomposition of the Lorentz group). The sections may be
treated in the same manner as the gauge fields so that they do not 
gain vevs. Consequently, new one loop topological contributions can 
arise in these situations as well.

\subsubsection*{Acknowledgements}

G.T.\ would like to thank M. Blau, E. Gava, K.S. Narain and S.
Randjbar-Daemi for helpful conversations. 

\rnc{\Large}{\normalsize}


\begin{thebibliography}{00}
\addcontentsline{toc}{section}{References}
\frenchspacing
\small
\addtolength{\itemsep}{-4pt}

\bibitem{red} A. Redlich, {\bf Parity Violation and Gauge Invariance of
the Effective Gauge Field Action in Three Dimensions}, Phys. Rev. D29
(1984) 2366.
\bibitem{seiberg5d} N. Seiberg, {\bf Five Dimensional SUSY Field
Theories, Non-Trivial Fixed Points and String Dynamics}, Phys. Lett.
388B (1996) 753, hep-th/9608111. 
\bibitem{seiberg3d} N. Seiberg, {\bf IR Dynamics on Branes and
Space-Time Geometry}, hep-th/9606017, Phys. Lett. 384B (1996)
81.
\bibitem{SW3d} N. Seiberg and E. Witten, {\bf Gauge Dynamics and
Compactification to Three Dimensions}, hep-th/9607163.
\bibitem{br} E. Bergshoeff and M. de Roo, {\bf D-Branes and T
Duality}, Phys. Lett. 380B (1996) 265, hep-th/9603123.
\bibitem{ght} M. Green, C. Hull and P. Townsend, {\bf D-Brane
Wess-Zumino Actions, T-Duality and the Cosmological Constant}, Phys.
Lett. 382B (1996) 65, hep-th/9604119.
\bibitem{dkmtv1} N. Dorey, V. Khoze, M. Mattis, D. Tong and S.
Vandoren, {\bf Instantons, Three-Dimensional Gauge Theory and the
Atiyah-Hitchin Manifold}, Nucl. Phys. B502 (1997) 59, 
hep-th/9703228.
\bibitem{CH3d} G. Chalmers and A. Hanany, {\bf Three Dimensional Gauge
Theories and Monopoles}, Nucl. Phys. B489 (1997) 223, 
hep-th/9608105.
\bibitem{Callias} C. Callias, Comm. Math. Phys. 62 (1978) 162.
\bibitem{sv1} V. Novikov, M. Shifman, A. Vainshtein, M. Voloshin and
V. Zakharov, Nucl. Phys. B229 (1983) 394.
\bibitem{sv2} M. A. Shifman and A. I. Vainshtein, Nucl. Phys. B277
(1986) 456.
\bibitem{sv3} M. A. Shifman and A. I. Vainshtein, Nucl. Phys. B359
(1991) 571.
\bibitem{sw4d} N. Seiberg and E. Witten, Nucl. Phys. B426 (1994) 19.
\bibitem{more5d} D. Morrison and N. Seiberg, {\bf Extremal Transitions
and Five-Dimensional Supersymmetric Field Theories}, Nucl. Phys. B483 
(1997) 229, hep-th/9609070. 
\bibitem{more5d2} K. Intriligator, D. Morrison and N. Seiberg, {\bf
Five-Dimensional Supersymmetric Gauge Theories and Degenerations of
Calabi-Yau Spaces}, Nucl. Phys. B497 (1997) 56, hep-th/9702198. 
\bibitem{witten5d} E. Witten, {\bf Phase Transitions in M-Theory and
F-Theory}, Nucl. Phys. B471 (1996) 195, hep-th/9603150.
\bibitem{bt} M. Blau and G. Thompson, {\bf Euclidean SYM Theories by
Time Reduction and Special Holonomy Manifolds}, hep-th/9706225,
to appear in Phys. Lett. B.
\bibitem{bsv} M. Bershadsky, V. Sadov and C. Vafa, {\bf D-Branes and
Topological Field Theories}, Nucl. Phys. B463 (1996) 166, hep-th/9511222.
\bibitem{bt2} M. Blau and G. Thompson, {\bf Aspects of $N_{T}\geq 2$
Topological Gauge Theories and D-Branes}, Nucl. Phys. B492 (1997) 545,
hep-th/9612143.
\bibitem{bbm} B.S. Acharya, M. O'Loughlin and B. Spence, {\bf Higher
Dimensional Analogues of Donaldson-Witten Theory}, hep-th/9705138.
\bibitem{bks}L. Baulieu, H. Kanno and I.M. Singer, {\bf Special
Quantum Field Theories in Eight and Other Dimensions}, 
hep-th/9704167.
\bibitem{bbjm}B. S. Acharya, J.M. Figueroa-O'Farrill, M. O'Loughlin 
and B. Spence, {\bf Euclidean D-Branes and Higher-Dimensional Gauge
Theory}, hep-th/9707118.
\end{thebibliography}
\end{document}